\documentclass[11pt,a4paper]{article}

\usepackage{epsfig}
\usepackage{amssymb}
\usepackage{amsmath}

\begin{document}

\title{Linearized stability of charged thin-shell wormholes}
\author{Ernesto F. Eiroa$^{1,}$\thanks{
e-mail: eiroa@iafe.uba.ar}, Gustavo E. Romero$^{2,}$\thanks{
Member of CONICET; e-mail: romero@irma.iar.unlp.edu.ar} \\
{\small $^1$ Instituto de Astronom\'{\i}a y F\'{\i}sica del Espacio, C.C.
67, Suc. 28, 1428, Buenos Aires, Argentina}\\
{\small $^2$Instituto Argentino de Radioastronom\'{\i}a, C.C.5, 1894 Villa
Elisa, Buenos Aires, Argentina}}

\maketitle

\begin{abstract}
The linearized stability of charged thin shell wormholes under spherically
symmetric perturbations is analyzed. It is shown that the presence of a 
large value of charge provides stabilization to the system, in the sense 
that the constraints onto the equation of state are less severe than for 
non-charged wormholes.\\

 
\end{abstract}

\section{Introduction}

Lorentzian wormholes were originally found as solutions of the Einstein
field equations with non-trivial topology. In their more simple examples
they present two mouths in asymptotically flat spacetimes connected by a
throat \cite{motho}. More general and purely geometric definitions allow 
to remove the requirement for asymptotically flat regions and the
introduction of trivial topologies in wormhole spacetimes, both static 
\cite{hovis1} and dynamic \cite{hovis2}.  All these wormholes, nonetheless,
must be threaded by matter that violates the null energy condition if they
are going to be traversable by material systems \cite{motho,hovis1,hovis2}.\\

The requirements for exotic matter that violates the energy conditions can 
be reduced introducing scalar fields or electric charges. Very recently 
Kim and Lee \cite{kim} have presented exact solutions for a static charged 
wormhole. These spacetimes are free of horizons, so the charge, although 
having effects on the stability, does not seriously affect the 
spacetime itself.\\   

Despite nothing is known about the equation of state of exotic matter, a
stability analysis is possible through linearized perturbations around
static solutions of the Einstein field equations. This imposes constraints
onto the state equation and the stability domain of the solutions. This
approach has been implemented by Poisson and Visser for spherically
symmetric thin-shell wormholes \cite{visser,poisson}. For surgical
techniques to construct thin-shell wormholes, the reader is referred to
the papers by Visser \cite{vis1, vis2}.\\

In this paper we extend the stability analysis to the case of charged 
wormholes with the aim to establish whether the presence of
charge effectively can increase the stability domain of this type of 
spacetimes. We do not intend to provide any explanation on the
mechanisms that might allow the wormhole to acquire or maintain its
charge, but we rather focus on the consequences of the charge on the
global stability of the system.

\section{Charged thin shell wormholes}

We use geometrized units (c=G=1). The Reissner-Nordstr\"{o}m metric is the
unique spherically symmetric solution of the vacuum Einstein-Maxwell coupled
equations. In Schwarzschild coordinates $X^{\alpha }=(t,r,\theta ,\varphi )$
it takes the form
\begin{equation}
ds^{2}=-\left( 1-\frac{2M}{r}+\frac{Q^{2}}{r^{2}}\right) dt^{2}+\left( 1- 
\frac{2M}{r}+\frac{Q^{2}}{r^{2}}\right) ^{-1}dr^{2}+r^{2}\left( d\theta
^{2}+\sin ^{2}\theta d\varphi ^{2}\right),  \label{e1}
\end{equation}
where $M$ is the mass and $Q^{2}$ is the sum of the squares of the electric 
($Q_{\rm{E}}$) and magnetic ($Q_{\rm{M}}$) charges, measured by distant 
observers. In a local orthonormal  frame, the nonzero components of 
the electromagnetic field tensor are 
$F^{\hat{t}\hat{r}}=E^{\hat{r}}=Q_{\rm{E}}/r^{2}$ and 
$F^{\hat{\theta }\hat{\varphi }}=B^{\hat{r}}=Q_{\rm{M}}/r^{2}$. If 
$\left| Q\right| \leq M$, this geometry represents a black hole with an event 
horizon with radius 
\begin{equation}
r_{h}=M+\sqrt{M^{2}-Q^{2}}.  \label{e2}
\end{equation}
If $\left| Q\right| >M$ it is a naked singularity.\\ 

From the Reissner-Nordstr\"{o}m geometry we can take two copies of the 
region with $r\geq a$:
\begin{equation}
\mathcal{M}^{\pm }=\{x/r\geq a\},  \label{e3}
\end{equation}
where $a>r_{h}$ if $\left| Q\right| \leq M$ and $a>0$ if $\left| Q\right|
>M, $ and glue them together at the hypersurface
\begin{equation}
\Sigma \equiv \Sigma ^{\pm }=\{x/r-a=0\},  \label{e4}
\end{equation}
to make a charged spherically symmetric thin shell wormhole $\mathcal{M}= 
\mathcal{M}^{+}\cup \mathcal{M}^{-}$. 
This construction creates a geodesically complete manifold $\mathcal{M}$ with 
two asymptotically plane regions connected by a throat. On $\mathcal{M}$ we 
can define a new radial coordinate $l=\pm \int _{a}^{r}g_{rr}dr$, where the 
positive and negative signs correspond, respectively, to $\mathcal{M}^{+}$ 
and $\mathcal{M}^{-}$. $|l|$ represents the proper radial distance to the 
throat, which is placed in $l=0$. Distant observers in $\mathcal{M}$ will see 
the wormhole as having mass $M$ and charges $Q_{\rm{E}}$ and $Q_{\rm{M}}$. To
study this traversable wormhole we use the standard Darmois-Israel formalism
\cite{daris}. For a modern treatment of this formalism see for example
\cite{mus}.\\

The wormhole throat is placed at $\Sigma $. This shell is a synchronous
timelike hypersurface. We can adopt coordinates $\xi ^{i}=(\tau ,\theta
,\varphi )$ in $\Sigma $, with $\tau $\ the proper time on the shell. In 
order to analyze the stability under spherically symmetric perturbations, 
we let the radius of the throat be a function of the proper time, 
$a=a(\tau )$. Then 
\begin{equation}
\Sigma :f(r,\tau )=r-a(\tau )=0.  \label{e5}
\end{equation}

The second fundamental forms associated with the two sides of the shell are: 
\begin{equation}
K_{ij}^{\pm }=-n_{\gamma }^{\pm }\left. \left( \frac{\partial ^{2}X^{\gamma
} } {\partial \xi ^{i}\partial \xi ^{j}}+\Gamma _{\alpha \beta }^{\gamma } 
\frac{ \partial X^{\alpha }}{\partial \xi ^{i}}\frac{\partial X^{\beta }}{
\partial \xi ^{j}}\right) \right| _{\Sigma },  \label{e6}
\end{equation}
where $n_{\gamma }^{\pm }$ are the unit normals ($n^{\gamma }n_{\gamma }=1$)
 to $\Sigma $ in $\mathcal{M}$: 
\begin{equation}
n_{\gamma }^{\pm }=\pm \left| g^{\alpha \beta }\frac{\partial f}{\partial
X^{\alpha }}\frac{\partial f}{\partial X^{\beta }}\right| ^{-1/2}\frac{
\partial f}{\partial X^{\gamma }}.  \label{e7}
\end{equation}
In the orthonormal basis $\{e_{\hat{\tau}},e_{\hat{\theta}},e_{\hat{\varphi}
}\}$ ($e_{\hat{\tau}}=e_{\tau }$, $e_{\hat{\theta}}=r^{-1}e_{\theta }$, $e_{
\hat{\varphi}}=(r\sin \theta )^{-1}e_{\varphi }$, $g_{_{\hat{\imath}\hat{
\jmath}}}=\eta _{_{\hat{\imath}\hat{\jmath}}}$) we have that 
\begin{equation}
K_{\hat{\theta}\hat{\theta}}^{\pm }=K_{\hat{\varphi}\hat{\varphi}}^{\pm
}=\pm \frac{1}{a}\sqrt{1-\frac{2M}{a}+\frac{Q^{2}}{a^{2}}+\dot{a}^{2}},
\label{e8}
\end{equation}
and 
\begin{equation}
K_{\hat{\tau}\hat{\tau}}^{\pm }=\mp \frac{\ddot{a}+\frac{M}{a^{2}}-\frac{
Q^{2}}{a^{3}}}{\sqrt{1-\frac{2M}{a}+\frac{Q^{2}}{a^{2}}+\dot{a}^{2}}},
\label{e9}
\end{equation}
where the dot means $d/d\tau $.\\

The Einstein equations on the shell reduce to the Lanczos equations: 
\begin{equation}
-[K_{\hat{\imath}\hat{\jmath}}]+Kg_{\hat{\imath}\hat{\jmath}}=8\pi S_{\hat{
\imath}\hat{\jmath}},  \label{e10}
\end{equation}
where $[K_{_{\hat{\imath}\hat{\jmath}}}]\equiv K_{_{\hat{\imath}\hat{\jmath}
}}^{+}-K_{_{\hat{\imath}\hat{\jmath}}}^{-}$, $K=tr[K_{\hat{\imath}\hat{
\jmath }}]=[K_{\hat{\imath}}^{\hat{\imath}}]$ and $S_{_{\hat{\imath}\hat{
\jmath} }}={\rm diag}(\sigma ,-\vartheta _{1},-\vartheta _{2})$ is the 
surface stress-energy tensor, with $\sigma $ the surface energy density and 
$\vartheta _{1,2}$ the surface tensions. Then replacing Eqs. (\ref{e8}) and 
(\ref{e9}) in Eq. (\ref{e10}) we obtain 
\begin{equation}
\sigma =\frac{-1}{2\pi a}\sqrt{1-\frac{2M}{a}+\frac{Q^{2}}{a^{2}}+\dot{a}
^{2},}  \label{e11}
\end{equation}
\begin{equation}
p=\frac{1}{4\pi a}\frac{1-\frac{M}{a}+\dot{a} ^{2}+a\ddot{a}}{\sqrt{1-\frac{
2M}{a}+\frac{Q^{2}}{a^{2}}+\dot{a}^{2}}}.  \label{e12}
\end{equation}
where $p=-\vartheta _{1}=-\vartheta _{2}$ is the surface pressure. The
surface energy density is negative, indicating the presence of 
\textit{exotic matter} in the throat.\\

The dynamical evolution of the wormhole throat is governed by the Einstein
equations plus the equation of state $p=p(\sigma )$ that relates $p$ and $
\sigma $. Replacing the equation of state in Eq. (\ref{e12}) and using Eq. 
(\ref{e11}), a second order differential equation is obtained for $a(\tau )$. 
This equation has an unique solution for given initial conditions 
$a(\tau _{0})$ and $\dot{a}(\tau _{0})$, where $\tau _{0}$ is an arbitrary 
(but fixed) time.\\ 

From the Eqs. (\ref{e11}) and (\ref{e12}) above, it is easy to verify the 
energy conservation equation: 
\begin{equation}
\frac{d}{d\tau }\left( \sigma A\right) +p\frac{dA}{d\tau }=0,  \label{e13}
\end{equation}
where $A=4\pi a^{2}$ is the area of the wormhole throat. The first term of
Eq. (\ref{e13}) represents the internal energy change of the throat and the
second the work done by the internal forces of the throat.
As in the case of uncharged wormholes \cite{poisson}, the conservation
equation can be written in the form 
\begin{equation}
\dot{\sigma}=-2(\sigma +p)\frac{\dot{a}}{a},  \label{e14}
\end{equation}
which can be integrated to give 
\begin{equation}
\ln \frac{a}{a(\tau _{0})}=-\frac{1}{2}\int ^{\sigma }_{\sigma (\tau _{0})}
\frac{d\sigma }{\sigma +p(\sigma )}.  \label{e15}
\end{equation}
This can be formally inverted to obtain $\sigma =\sigma (a)$. Thus, 
replacing $
\sigma (a)$ in Eq. (\ref{e11}) and regrouping terms, the dynamics of the
wormhole throat is completely determined by a single equation: 
\begin{equation}
\dot{a}^{2}-\frac{2M}{a}+\frac{Q^{2}}{a^{2}}-4\pi ^{2}a^{2}\left[ \sigma (a) 
\right] ^{2}=-1,  \label{e16}
\end{equation}
with the initial conditions $a(\tau _{0})$ and $\dot{a}(\tau _{0})$.

\section{Stability}

\begin{figure}[t!]
\vspace{-5cm} 
\includegraphics[width=16cm,height=22.5cm]{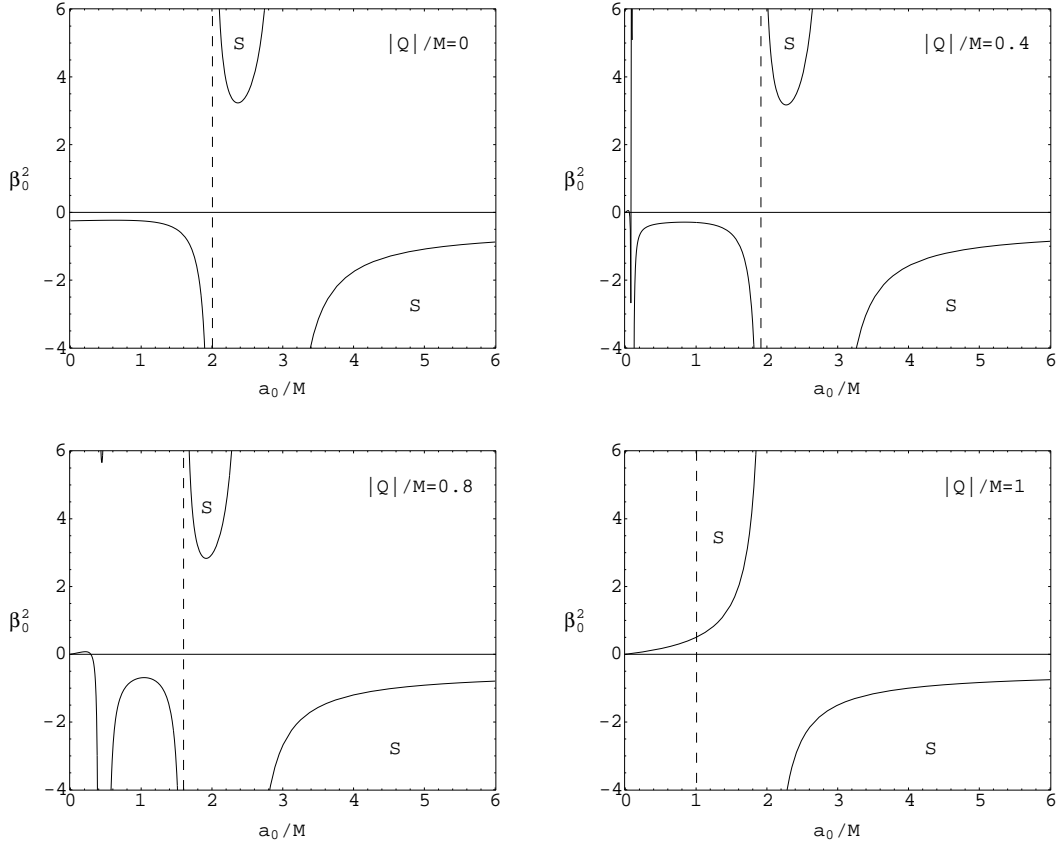} 
\vspace{-6cm}
\caption{Stability regions (S) for different values of charge with $
|Q|/M\leq 1 $. The zones situated at the left of the dashed vertical lines
have no physical meaning (they correspond to $a_{0}\leq r_{h}$).}
\label{fig1}
\end{figure}

\begin{figure}[t!]
\vspace{-5cm} 
\includegraphics[width=16cm,height=22.5cm]{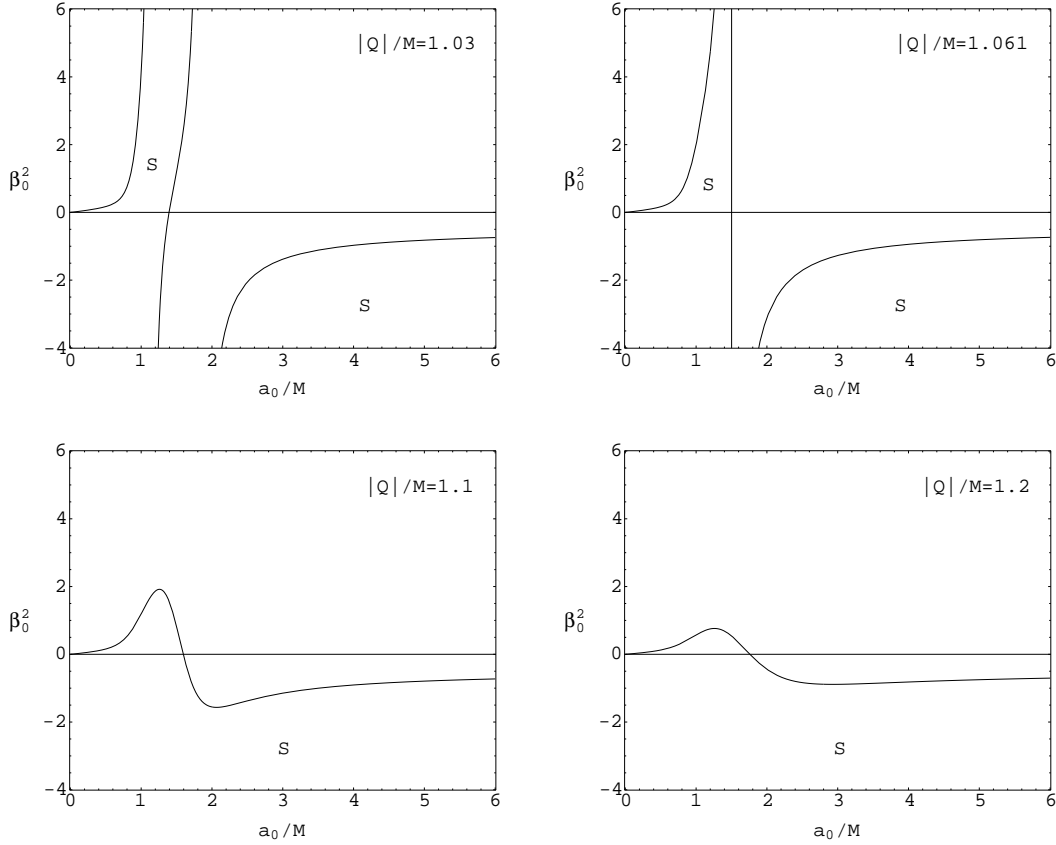} 
\vspace{-6cm}
\caption{Stability regions (S) for different values of charge with $|Q|/M>1 $
.}
\label{fig2}
\end{figure}

Eq. (\ref{e16}) can be rewritten in the form: 
\begin{equation}
\dot{a}^{2}=-V(a),  \label{e17}
\end{equation}
with 
\begin{equation}
V(a)=1-\frac{2M}{a}+\frac{Q^{2}}{a^{2}}-4\pi ^{2}a^{2}[\sigma (a)]^{2}.
\label{e18}
\end{equation}

Let us consider now a static solution with radius $a_{0}$ and the 
corresponding values of energy density $\sigma _{0}$ and pressure $p_{0}$: 
\begin{equation}
\sigma _{0}=\frac{-1}{2\pi a_{0}}\sqrt{1-\frac{2M}{a_{0}}+\frac{Q^{2}}{
a_{0}^{2}},}  \label{e19}
\end{equation}
\begin{equation}
p_{0}=\frac{1}{4\pi a_{0}}\frac{1-\frac{M}{a_{0}}}{\sqrt{1-\frac{2M}{a_{0}}+ 
\frac{Q^{2}}{a_{0}^{2}}}}.  \label{e20}
\end{equation}
To study the stability of this static solution under radial perturbations, we 
carry out a linearized analysis, as it is done for the noncharged case in 
\cite{poisson}. A Taylor expansion to second order of the potential $V(a)$ 
around the static solution yields: 
\begin{equation}
V(a)=V(a_{0})+V^{\prime }(a_{0})(a-a_{0})+\frac{V^{\prime \prime }(a_{0})}{2}
(a-a_{0})^{2}+O(a-a_{0})^{3},  
\label{e21}
\end{equation}
where the prime means $\frac{d}{da}$. From Eq. (\ref{e18}) the first
derivative of $V(a)$ is 
\begin{equation}
V^{\prime }(a)=\frac{2M}{a^{2}}-\frac{2Q^{2}}{a^{3}}-8\pi ^{2}a\sigma
(\sigma +a\sigma ^{\prime }).  \label{e22}
\end{equation}
Since $\sigma ^{\prime }=\dot{\sigma}/\dot{a}$, using the Eq. (\ref{e14}), 
we have that $\sigma ^{\prime }=-2(\sigma +p)/a$. Then 
\begin{equation}
V^{\prime }(a)=\frac{2M}{a^{2}}-\frac{2Q^{2}}{a^{3}}+8\pi ^{2}a\sigma
(\sigma +2p).  \label{e23}
\end{equation}
The second derivative of the potential is 
\begin{equation}
V^{\prime \prime }(a)=-\frac{4M}{a^{3}}+\frac{6Q^{2}}{a^{4}}+8\pi ^{2}\left[
(\sigma +a\sigma ^{\prime })(\sigma +2p)+a\sigma (\sigma ^{\prime
}+2p^{\prime })\right] .  
\label{e24}
\end{equation}
Since $\sigma ^{\prime }+2p^{\prime }=\sigma ^{\prime }(1+2p^{\prime }/\sigma
^{\prime })$, defining the parameter $\beta ^{2}(\sigma )\equiv dp/d\sigma
=p^{\prime }/\sigma ^{\prime }$, we have that $\sigma ^{\prime }+2p^{\prime
}=\sigma ^{\prime }(1+2\beta ^{2})$. Using again that $\sigma ^{\prime
}=-2(\sigma +p)/a$, we obtain 
\begin{equation}
V^{\prime \prime }(a)=-\frac{4M}{a^{3}}+\frac{6Q^{2}}{a^{4}}-8\pi ^{2}\left[
(\sigma +2p)^{2}+2\sigma (\sigma +p)(1+2\beta ^{2})\right] .  
\label{e25}
\end{equation}
Replacing in $a=a_{0}$, we have that $V(a_{0})=V^{\prime }(a_{0})=0$, so
\begin{equation}
V(a)=\frac{1}{2}V^{\prime \prime }(a_{0})(a-a_{0})^{2}+O[(a-a_{0})^{3}],
\label{e26}
\end{equation}
with 
\begin{equation}
V^{\prime \prime }(a_{0})=\frac{-2}{a_{0}^{4}}\left[ \frac{a_{0}\left[ 
(a_{0}-M)^{3}+M(M^{2}-Q^{2})\right] }{a_{0}^{2}-2Ma_{0}+Q^{2}}
+2\left( a_{0}^{2}-3Ma_{0}+2Q^{2}\right) \beta _{0}^{2}\right] ,  
\label{e27}
\end{equation}
where $\beta _{0}=\beta (\sigma _{0})$.\\

The wormhole is stable if and only if $V^{\prime \prime }(a_{0})>0$. The
curves where $V^{\prime \prime }(a_{0})=0$ correspond to 
\begin{equation}
\beta _{0}^{2}=f_{0}(a_{0})\equiv \frac{-a_{0}\left[ (a_{0}-M)^{3}+
M(M^{2}-Q^{2})\right] }{2\left( a_{0}^{2}-2Ma_{0}+Q^{2}\right) \left(
a_{0}^{2}-3Ma_{0} +2Q^{2}\right) },  
\label{e28}
\end{equation}
and $a_{0}=3M/2$ $(\beta _{0}^{2}\in \mathbb{R})$ when $|Q|=3M/\sqrt{8}$.\\

For the analysis of the stability regions it is convenient to consider five
cases accordingly to the value of charge:
\begin{enumerate}
\item  Case $0\leq \frac{|Q|}{M}<1$. There are two regions of stability,\\

i) $\beta _{0}^{2}>f_{0}(a_{0})$ if $1+\sqrt{1-\frac{Q^{2}}{M^{2}}}<\frac{
a_{0}}{M}<\frac{3}{2}+\frac{1}{2}\sqrt{9-8\frac{Q^{2}}{M^{2}}}$, and\\

ii) $\beta _{0}^{2}<f_{0}(a_{0})$ if $\frac{a_{0}}{M}>\frac{3}{2}+\frac{1}{2}
\sqrt{9-8\frac{Q^{2}}{M^{2}}}$.
\item  Case $\frac{|Q|}{M}=1$. There are two regions of stability,\\

i) $\beta _{0}^{2}>f_{0}(a_{0})$ if $1<\frac{a_{0}}{M}<2$, and\\

ii) $\beta _{0}^{2}<f_{0}(a_{0})$ if $\frac{a_{0}}{M}>2$.
\item  Case $1<\frac{|Q|}{M}<\frac{3}{\sqrt{8}}$. There are two regions of
stability,\\

i) $\left\{ 
\begin{array}{ll}
\beta _{0}^{2}<f_{0}(a_{0}) & 
\mbox{if $0< \frac{a_{0}}{M}<\frac{3}{2}-\frac{1}{2}
\sqrt{9-8\frac{Q^{2}}{M^{2}}}$} \\ 

\beta _{0}^{2}\in \mathbb{R} & 
\mbox{if $\frac{a_{0}}{M}=\frac{3}{2}-\frac{1}{2}
\sqrt{9-8\frac{Q^{2}}{M^{2}}}$} \\ 

\beta _{0}^{2}>f_{0}(a_{0}) & 
\mbox{if $\frac{3}{2}-\frac{1}{2}
\sqrt{9-8\frac{Q^{2}}{M^{2}}}<\frac{a_{0}}{M}<\frac{3}{2}+\frac{1}{2}
\sqrt{9-8\frac{Q^{2}}{M^{2}}}$}
\end{array}
\right. $\\

and\\

ii) $\beta _{0}^{2}<f_{0}(a_{0})$ if $\frac{a_{0}}{M}>\frac{3}{2}+\frac{1}{2}
\sqrt{9-8\frac{Q^{2}}{M^{2}}}$.
\item  Case $\frac{|Q|}{M}=\frac{3}{\sqrt{8}}$. There are two regions of
stability,\\

i) $\beta _{0}^{2}<f_{0}(a_{0})$ if $0<\frac{a_{0}}{M}<\frac{3}{2}$, and\\

ii) $\beta _{0}^{2}<f_{0}(a_{0})$ if $\frac{a_{0}}{M}>\frac{3}{2}$.
\item  Case $\frac{|Q|}{M}>\frac{3}{\sqrt{8}}$. There is one region of
stability,\\

$\beta _{0}^{2}<f_{0}(a_{0})$ if $\frac{a_{0}}{M}>0$.\\
\end{enumerate}

The regions of stability for different values of charge are shown in Figs. 
\ref{fig1} and \ref{fig2}.
For ordinary matter, $\beta _{0}$ represents the velocity of sound, so it
should satisfy $0< \beta _{0}^{2}\leq 1$. If the matter is exotic, as it
happens to be in the throat,  $\beta _{0}$ is not necessarily the velocity of
sound and it is not clear which values it can take (see discussion in \cite
{poisson}).\\

For $0<|Q|/M\leq 1$ the stability regions are similar to the case of uncharged
wormholes. The \textit{tongue} shaped region with $\beta _{0}^{2}>0$ moves
slowly to lower values of $\beta _{0}^{2}$ and $a_{0}/M$ as the charge
increases. Values of  $\beta _{0}^{2}$ in the interval $(0,1]$ are not
included in the stability region unless $|Q|/M$ is very close to one. The
other region of stability only moves slowly to lower values of $a_{0}/M$ for
higher values of charge.\\

The situation changes dramatically if $1<|Q|/M\leq 3/\sqrt{8}$. In this
case, for every value of the parameter  $\beta _{0}^{2}$ there are values of
the radius $a_{0}/M$ that makes the wormhole stable. When $1<\frac{|Q|}{M}<
\frac{3}{\sqrt{8}}$ and $\frac{a_{0}}{M}=\frac{3}{2}-\frac{1}{2}
\sqrt{9-8\frac{Q^{2}}{M^{2}}}$, it is stable regardless of the equation of 
state of the exotic matter (which determines $\beta _{0}^{2}$).
If $|Q|/M>3/\sqrt{8}$, besides the large negative $\beta _{0}^{2}$ region of
stability, it includes also a small one with positive $\beta _{0}^{2}$ ,
for radii of the throat in the range $0<\frac{a_{0}}{M}<1+\sqrt[3]{\frac{
Q^{2}}{M^{2}}-1}$.

\section{Conclusions}

We have applied a linearization stability analysis to thin-shell wormholes 
endowed with charge and we have found that the presence of large charges 
($|Q|/M \approx 1$), significantly increases the possibility of obtaining 
stable wormhole spacetimes, imposing less severe restrictions on the equation 
of state of the exotic matter that must be present in the wormhole throat. 
Although it is not possible to give a full physical meaning to the parameter 
$\beta_0$ in the absence of a detailed microphysical model for the exotic 
matter, our analysis shows that, for a given radius $a_{0}$, it is {\em always} 
possible to find stable traversable wormhole solutions for {\em any} value of 
$\beta_0$ by taking adequate values of charge and mass. 


\section*{Acknowledgments}

This work has been supported by Universidad de Buenos Aires (UBACYT X-143,
EFE), CONICET (PIP 0430/98, GER), ANPCT (PICT 98 No. 03-04881, GER), and
Fundaci\'{o}n Antorchas (GER). GER thanks the Max-Planck-Institut 
f\"ur Kernphysik for kind hospitality and Dr. Diego F. Torres for 
discussions. Some calculations in this paper 
were done with the help of the package GRTensorII \cite{grt}.

\end{document}